\documentclass[%
 reprint,
 amsmath,amssymb,
 superscriptaddress,
 aps,
prl,
]{revtex4-2}

\usepackage{graphicx}
\usepackage{blindtext}
\usepackage{dcolumn}
\usepackage{bm}
\usepackage{braket}
\usepackage{setspace}
\usepackage{color}

\usepackage[english]{babel}

\begin{document}
\preprint{APS/123-QED}

\title{Universal scaling hypothesis of quantum spatial search in complex networks}

\author{Rei Sato}
\email{reisato\_0018@icloud.com}
\affiliation{%
  Department of Physics, Tokyo University of Science, Shinjuku, Tokyo, 162-8601, Japan\\
}%
\affiliation{%
 Department of Transdisciplinary Science and Engineering, School of Environment and Society, Tokyo Institute of Technology,
2-12-1 O-okayama, Meguro-ku, Tokyo 152-8550, Japan\\
}%
\author{Tetsuro Nikuni}%
\affiliation{%
 Department of Physics, Tokyo University of Science,
 Shinjuku, Tokyo, 162-8601, Japan\\
}%

\author{Kayoko Nohara}%
\affiliation{%
 Department of Transdisciplinary Science and Engineering, School of Environment and Society, Tokyo Institute of Technology,
2-12-1 O-okayama, Meguro-ku, Tokyo 152-8550, Japan\\
}%

\author{Giorgio Salani}%
\affiliation{%
 Department of Transdisciplinary Science and Engineering, School of Environment and Society, Tokyo Institute of Technology,
2-12-1 O-okayama, Meguro-ku, Tokyo 152-8550, Japan\\
}%

\author{Shohei Watabe}%
\email{watabe@shibaura-it.ac.jp}
\affiliation{%
  Department of Physics, Tokyo University of Science, Shinjuku, Tokyo, 162-8601, Japan\\
}%
\affiliation{%
   Faculty of Engineering, Computer Science and Engineering, Shibaura Institute of Technology, Toyosu, Tokyo, 135-8548, Japan\\
}%

\begin{abstract} 
Since quantum spatial searches on networks have a strong network dependence, the question arises whether the universal perspective exists in this quantum algorithm for various networks. 
Here, we propose a universal scaling hypothesis of the quantum spatial search on complex networks such as small-world and scale-free networks. 
The average path length, a key quantity in the complex network science, is useful to expose this universal feature, where the collapse plot can be generated for the optimal time, the maximal finding probability and the optimal hopping parameter. 
Based on the path integral method, we also clarify that the probability amplitude in the continuous-time quantum walk can be determined by the path length distribution. 
\end{abstract}

\maketitle


Network science is a cornucopia of nontrivial properties covering a wide range of fields, such as classical physics, society, biology and artificial intelligence~\cite{balabasi1999,watts1998collective,song2005self,amaral2004complex,newman2003structure,RevModPhys.74.47,barabasi2016network,Newmanbook}. 
Indeed, complex network science is helpful to understand real societies, for example, community detection~\cite{Zachary1977,Reichardt:2004ea}, cascade control~\cite{Motter:2004io}, and contagion phenomena~\cite{doi:10.1126/science.1245200}. 
This network science is a fountain of nontrivial concepts behind complexity. 
In particular, classical network science not only provides the concept of the structure of networks, such as random~\cite{Erdos2022OnRG,10.1103/physreve.64.026118}, small-world~\cite{watts1998collective,newman1999scaling}, scale-free~\cite{balabasi1999,10.1103/physreve.64.046135} and self-similar~\cite{10.1103/physrevlett.96.018701,song2005self,PhysRevE.72.045105} networks, but also reveals dynamical properties 
~\cite{10.1103/physreve.69.066113,10.1103/physreve.72.056119,hwang2010spectral, 10.1038/ncomms6121,Barzel0w8}.  

However, the scope of the network science is now intimately involved in the quantum world~\cite{Biamonte2019}, such as quantum random networks~\cite{perseguers2010quantum}, 
quantum sensing networks~\cite{Guo:2020el,Rubio2020}, 
quantum communication~\cite{Hahn:2019ei}, 
entanglement percolation~\cite{Acin:2007im,Cuquet:2009il}, 
and 
quantum internet~\cite{Kimble:2008if,Simon:2017es,Wehner:2018cu,PhysRevLett.124.210501}. 
Interestingly, notions of the classical complex network science have also been applied to appreciate quantum systems, such as quantum phase transitions~\cite{Valdez2017}, time crystals~\cite{doi:10.1126/sciadv.aay8892}, and localization/delocalization~\cite{Jahnke2008,Bueno2020}.

Tools to analyze the complex networks are also multi-disciplined ranging from the graph theory~\cite{barabasi2016network,Newmanbook} to the percolation~\cite{newman1999scaling}, Bose--Einstein condensation~\cite{Bianconi:2001iq}, random matrix~\cite{PhysRevE.76.026109}, renormalization~\cite{PhysRevLett.101.148701}, quantum annealing~\cite{Negre:2020ev}, and quantum walks~\cite{shenvi2003quantum,Faccin:2014cda,PhysRevResearch.2.023378,Dernbach:2019jw,PhysRevA.107.032605,9828952,SanchezBurillo:2012gf,PhysRevA.96.032305,Wang2022,wrd,Whitfield:2010ki,Faccin:2013gq}. 
Indeed, the quantum walk, which has been recently demonstrated in experiments~\cite{10.1103/physrevlett.104.100503,10.1103/physrevlett.104.050502,Xiao:2018ita,doi:10.1126/science.1174436}, is a method widely used in the complex network science, such as the community detection~\cite{Faccin:2014cda,PhysRevResearch.2.023378}, link prediction~\cite{PhysRevA.107.032605,9828952}, centrality testing~\cite{PhysRevA.96.032305,Wang2022}, an element ranking~\cite{SanchezBurillo:2012gf} and neural networks~\cite{Dernbach:2019jw}.  

The combination of quantum walks and network science has highlighted their potential applications in various domains, in particular in machine learning area, where the performance of the quantum walks on graphs has been shown to be equal to or better than that of the classical random walks~\cite{dernbach2019quantum, 10.1007/978-981-97-2242-6_8}. 
Node classification, for instance, which involves labels or categories of nodes, has been extensively explored by classical algorithms~\cite{kipf2016semi, schlichtkrull2018modeling, perozzi2014deepwalk,grover2016node2vec}. 
Random-walk sampling algorithms on scale-free networks 
reduce learning costs of graph~\cite{perozzi2014deepwalk, grover2016node2vec}. 
The quantum walks, as the quantum counterpart of random walks, have also been investigated for learning graph representations.  Recent studies suggest that quantum walks can learn these representations more accurately than classical methods~\cite{10.1007/978-981-97-2242-6_8}.  However, the time complexity of spatial search on networks, such as scale-free and small-world networks, requires further investigation. This raises the question of whether a universal perspective of quantum dynamics exists within these algorithms for complex networks, potentially offering crucial insights into quantum systems and algorithms.

In particular, one of the central issues in the quantum spatial search is to reveal the relation between the quantum algorithm and the network structure, such as the scaling relation between the optimal number of oracle calls $Q$ and the number of nodes $N$, where the scaling $Q = {\mathcal O} (N^{1/2})$ holds in the complete graph~\cite{childs2004spatial} and conditionally in the Erd\"{o}s-R\'enyi random graph~\cite{chakraborty2016spatial}, 
whereas the scaling depends on the spectral dimension in the fractal lattices~\cite{10.7566/jpsj.87.085003,PhysRevA.86.012332,sato2020scaling} and on the closeness centrality in the scale-free network~\cite{osada2020continuous}. 
In particular, the quantum spatial search using the quantum walk has been extensively investigated for various network structures, {\it e.g.}, regular lattices~\cite{wrd,childs2004spatial,PhysRevA.82.032330,PhysRevA.82.032331}, random graphs~\cite{chakraborty2016spatial,Glos2018,Osada:2018ej}, fractal lattices~\cite{10.1103/physreve.90.032113,agliari2010quantum,10.7566/jpsj.87.085003,sato2020scaling,PhysRevA.98.012320,PhysRevA.86.012332}, and complex networks~\cite{D.I.Tsomokos,osada2020continuous}. 
Because of the complexity, however, the scaling of the optimal number of oracle calls $Q$ depends on the structure of the complex network.

Here we uncover universal properties behind the quantum spatial search in complex networks. 
The scaling with respect to the number of nodes shows the network dependence, which indicates that such a conventional analysis is useless to obtain the universal perspective. 
However, the average path length, which is a key quantity in the complex network science, is found to be crucial in generating collapse plots for various kinds of complex networks, such as small-world networks, the scale-free networks, as well as random, ring and fractal networks. 
Furthermore, we unveiled a class of networks, including not only the complex network model, such as the Watts--Strogatz (WS) model, but also real-world networks, {\it e.g.}, co-authorship as well as human disease and mouse brain networks, where the finding probability and the optimal hopping parameter are correlated and those data can be collapsed.  
Our results provide a new insight into the connection between quantum physics and complex network science, and will facilitate further integration of these fields.

\begin{figure}[tbp]
  \centering
    \includegraphics[width=80mm]{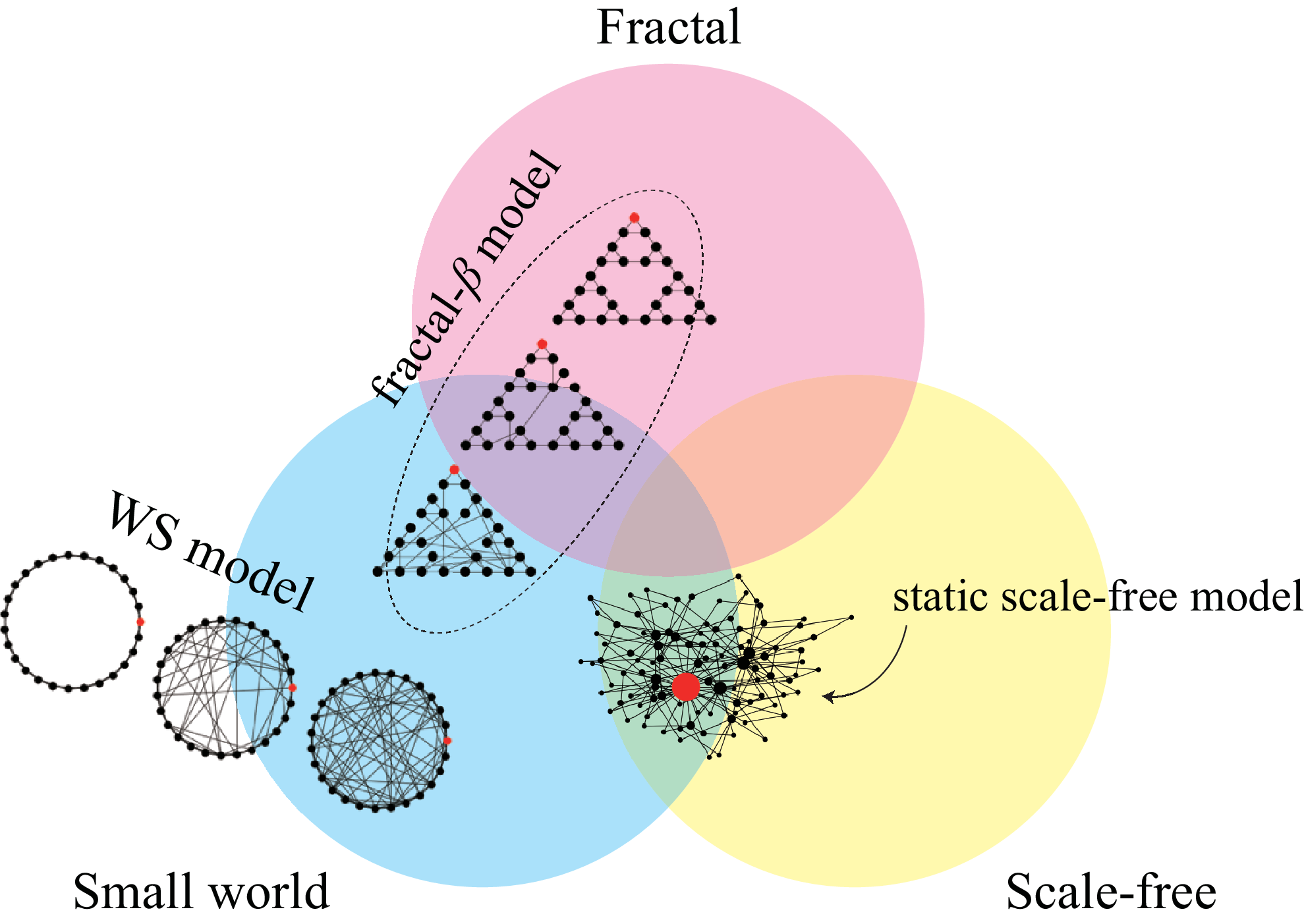}
    \caption{(Color online) A diagram of three characteristic structures of complex networks in this study: small-world, scale-free and fractal networks, which can be covered by the Watts--Strogatz (WS) model, static scale-free model and fractal-$\beta$ model. }
    \label{fig:complex_networks}
\end{figure}

To tackle the problem of unveiling the universality of the quantum spatial search in complex networks, 
we need to prepare various kinds of complex networks. 
We therefore utilize three models: the WS model~\cite{watts1998collective}, the fractal-$\beta$ model based on the Sierpinski gasket~\cite{tomochi2010model}, and the static scale-free network model~\cite{Goh2001} (Fig.~\ref{fig:complex_networks}). 
With these models and their variant models with weighted graphs, we can cover networks ranging from regular to random, fractal, scale-free, and small-world networks, where we can prepare these unweighted and weighted graphs. 
In the WS model~\cite{watts1998collective} and fractal-$\beta$ model~\cite{tomochi2010model}, 
the network structure is controlled by the parameter $\beta$, 
where the regular graph emerges at $\beta = 0$, the random graph at $\beta = 1$, 
and the small-world network at the intermediate value of $\beta$. 
The regular graphs in the WS and fractal-$\beta$ models show the ring network and the Sierpinski gasket, respectively. 
For the static scale-free network model, the network is characterized by the power-law degree distribution $P_k \sim k^{-\lambda}$, 
where an anomalous regime with a hub-and-spoke structure at $\lambda<2$, the scale-free regime with an ultra-small-world property at $2<\lambda<3$, 
and a random regime with a small-world property at $\lambda>3$. 
At $\lambda = 2,3$, the critical points emerge.

The continuous-time quantum walk on an unweighted graph $G(V,E)$ and weighted graph $G(V,E,W)$ takes place in an $N$-dimensional Hilbert space enclosed by states $\ket{i}$ with a node label $i=1,2,...,N$~\cite{childs2004spatial}.  

Here, the graph $G$ is a connected and undirected network with $N$-vertices and no self-loop. 
The quantum spatial search is achieved by utilizing the Schr\"odinger dynamics with the time-independent Hamiltonian $\hat{H} = \hat{H}_{L} + \hat{H}_w$. 
Here, $\hat{H}_w$ is the oracle term given by $\hat{H}_w \equiv - \gamma_w \ket{w}\bra{w}$, where $w$ is a target node, and $\gamma_w$ is a positive real number, which is normalized in this study, i.e., $\gamma_w = 1$. 
The diffusion (or hopping) term $\hat{H}_{L}$ is given by $\hat{H}_{L}\equiv \gamma \hat{L}$ with a positive real number $\gamma$ and a weighted-Laplacian operator $\hat{L} \equiv \hat{D} - \hat{A}$~\cite{Xu_2020}. 
Here, $\hat{A}$ is the weighted adjacency operator given by $\hat{A} \equiv \sum_{i,j} A_{ij} \ket{i}\bra{j}$, and $\hat{D}$ is a weighted node-degree operator given by $\hat{D} \equiv \sum_i d_i \ket{i}\bra{i}$ with the weighted node-degree $d_i$ that is the sum of the edge weights connected to the node $i$, i.e., $d_i \equiv \sum_j A_{ij}$.  
For the quantum search for the unweighted graph, we employ $A_{ij} = 0$ or $1$ for all the edge weights~\cite{childs2004spatial}, the value of which depends on the network structure we are interested in. 
For the weighted graph in this study, we replace the weight $A_{ij}=1$ in the unweighted graph with the number randomly drawn from the uniform distribution over the interval $(0,1]$.

An initial state for the quantum spatial search is a uniform superposition state, given by $\ket{s}= \sum_{i=1}^{N}\ket{i}/\sqrt{N}$. 
The finding probability $P(t)$ at a time $t$ on a target node $w$ is $P(t)=|\bra{w}e^{-i\hat{H}t}\ket{s}|^2$. 
In the quantum spatial search, the hopping parameter $\gamma$ is optimized so that the peak value of the finding probability is maximized~\cite{childs2004spatial}. 
This condition well corresponds to the condition where the overlap between $|s\rangle$ and the ground state $|E_0\rangle$ of $\hat H$ is equal to that for the first-excited state $|E_1 \rangle$~\cite{childs2004spatial}, {\rm i.e.,} $|\braket{s|E_0}|^2 = |\braket{s|E_1}|^2 \sim 0.5$. 
(Hereafter, $\gamma$ is denoted as the optimal value for simplicity.)
Using this optimal $\gamma$, we determine the optimal finding probability $P$ as well as the optimal time $t=Q$, which gives the peak value of the finding probability and its peak-to-peak period, respectively.

\begin{figure}[tbp]
    \centering
    \includegraphics[width=85mm]{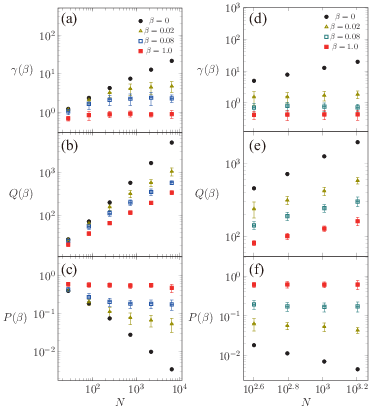}
    \caption{(Color online) Conventional plots of the optimal hopping parameter $\gamma$, the optimal time $Q$ and the optimal finding probability $P$ as a function of the number of nodes $N$ in the unweighted fractal-$\beta$ model (a)-(c), and in the unweighted WS model (d)-(f), respectively. 
    }
    \label{fig:N_scaling_smallworld}
\end{figure}


\begin{figure}[tbp]
    \centering
    \includegraphics[width=85mm]{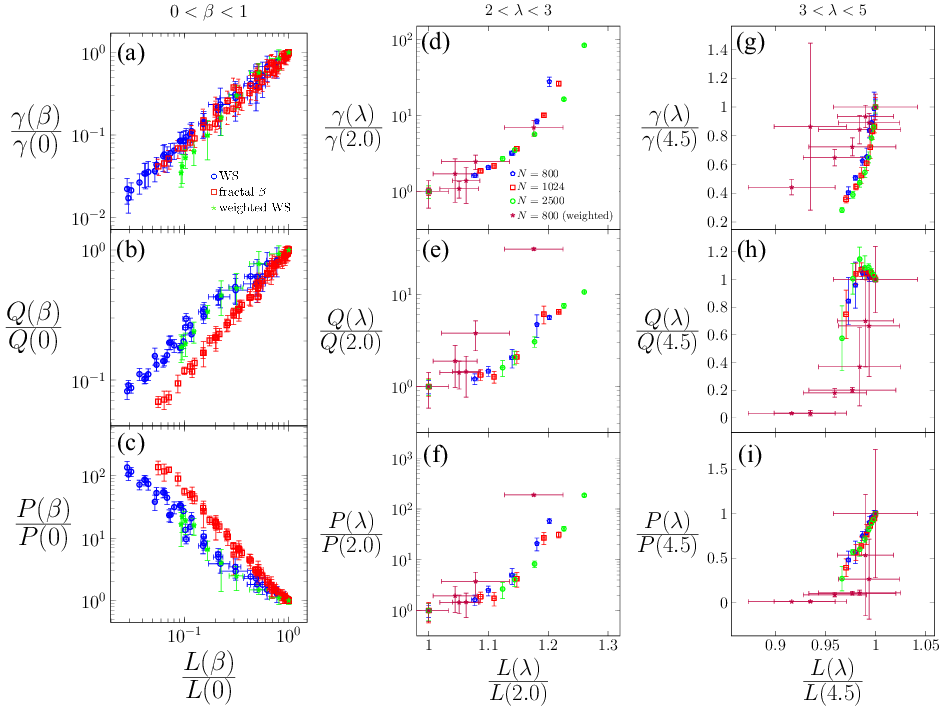}
    \caption{(Color online) Path-length dependence of $\gamma$, $Q$ and $P$ in the WS model and fractal-$\beta$ model (a)-(c), where all the quantities are normalized with each quantity of the regular network ($\beta = 0$) with the same number of nodes $N$. 
    Figures (d)-(f) are for the static scale-free model in the scale-free regime $(2 < \lambda < 3)$, and (g)-(i) are in the random regime $(3 < \lambda)$, where a target node is a hub (a node with the maximum degree) and quantities are normalized with those at $\lambda = 2$ and $4.5$, respectively. 
     }
   \label{fig:path_norm_scaling}
\end{figure}

Universal perspectives cannot be obtained if the conventional analysis of the quantum search algorithm is employed, where the scaling relations are supposed as $\gamma=\mathcal{O}(N^{a_{\gamma}})$, $Q=\mathcal{O}(N^{a_{Q}})$, and $P=\mathcal{O}(N^{a_{P}})$ (Fig.~\ref{fig:N_scaling_smallworld}). 
The network shows the crossover from the regular to random graphs through the small-world network in the WS model as well as the fractal-$\beta$ model, where we used unweighted graphs in Fig.~\ref{fig:N_scaling_smallworld}. 
The exponents $\alpha_{\gamma, Q, P}$ depend on the parameter $\beta$ as well as the network models, which is also the case for the static scale-free model with the parameter $\lambda$ giving the anomalous, scale-free and random regimes. 
Moreover, in the fractal-$\beta$ model, strictly speaking, $\gamma$, $Q$ and $P$ are not exactly a power function of $N$ except for $\beta = 0$ (Figs.~\ref{fig:N_scaling_smallworld} 
 (a)-(c)).

Here we seek a universal perspective applicable to complex networks. 
The quantum spatial search reflects the global structures of the complex network, because the initial state is the equal superposition state $|s\rangle$, and the wave function diffuses over the complex network throughout the quantum search process. 
It would be interesting to employ a length scale that characterizes the entire system, inspired by the context of the universality in the continuous phase transition in the statistical physics~\cite{newmanb99}. 
A possible characteristic length is an average path length $L$, rather than a path length between two specific nodes.

Figure~\ref{fig:path_norm_scaling} shows the average path length dependence of $\gamma$, $Q$ and $P$, where the data for various numbers of sites and various values of parameters $\beta$ and $\lambda$ are integrated. 
In Figs.~\ref{fig:path_norm_scaling} (a)-(c), the data for the WS and fractal-$\beta$ models were normalized with data for the regular network ($\beta=0$) with the same number of nodes $N$. 
For the static scale-free model (Figs.~\ref{fig:path_norm_scaling} (d)-(i)), the data were normalized with data for a specific value of $\lambda$. By using the normalization, we can surprisingly and successfully generate collapse plots, particularly in unweighted graphs. 
We have also tried to make similar plots as a function of the average clustering coefficient, for which we could not find any collapse plots in contrast to the case for the average path length. 
As a result, the average path length, not the average clustering coefficient, is a key to exposing the universality of the quantum spatial search in the complex network. 
For the weighted WS model, the data successfully collapse into those in the unweighted WS model (Figs.~\ref{fig:path_norm_scaling} (a)-(c)). 
For the scale-free regime at $2<\lambda<3$ (Figs.~\ref{fig:path_norm_scaling} (d)-(f)), 
the data structure is almost the same for the weighted and unweighted graphs, while the collapse plot does not work in a random regime at $\lambda>3$ (Figs.~\ref{fig:path_norm_scaling} (g)-(i)). 

The collapse plots for the small-world models have the model dependence (Figs.~\ref{fig:path_norm_scaling} (a)-(c)).
The model-specific behavior emerges around $\beta = 0$, where the regular lattice structure, such as the ring network and the fractal network, is dominant. 
In the small-world network regime and random network regime, however, 
the exponents are common between the unweighted and weighted WS models and the fractal-$\beta$ model. 
Indeed, we have the following scaling law for the WS model and fractal-$\beta$ model, given by 
\begin{align}
\frac{\gamma(\beta)}{\gamma(0)} &= \left(\frac{L(\beta)}{L(0)}\right)^{\alpha_\gamma}, 
\label{eq:ws_scaling_law_gamma}\\
\frac{Q(\beta)}{Q(0)} &= \left( \frac{L(\beta)}{L(0)}\right)^{\alpha_Q}, 
\label{eq:ws_scaling_law_Q}\\
\frac{P(\beta)}{P(0)} &= \left( \frac{L(\beta)}{L(0)}\right)^{\alpha_P}, 
\label{eq:ws_scaling_law_P}
\end{align}
where $\alpha_\gamma = 1.08(6)$, $\alpha_Q = 0.90(5)$, and $\alpha_P = -1.76(8)$. 

The average-path-length plot is still efficient for generating the collapse plots in the scale-free network, whereas the $L$-dependence is completely different in the scale-free regime (Figs.~\ref{fig:path_norm_scaling} (d)-(f) for $2<\lambda <3$) and the random regime (Figs.~\ref{fig:path_norm_scaling} (g)-(i) for $\lambda > 3$). 
The average path length dependence clearly emerges in $\gamma$, $Q$ and $P$ if the target is a hub, 
whereas its dependence is weak or absent for a target with a small degree. 
Although the average-path-length dependence is completely different between the scale-free network and the small-world network, we can surprisingly generate the collapse plot by utilizing the average path length in both networks. 
Since the results with various network sizes are plotted on a single curve, 
our finding of universality will be useful in predicting the behavior of quantum networks that cannot be simulated in classical computers by employing a small-size network that is tractable in classical computers.

The rigorous mathematical proof of the path-length scaling law is unfortunately still open. Therefore, our universal scaling remains a hypothesis to be proved analytically. 
However, a sensible physical explanation behind the relation between the quantum spatial search and the path length in complex networks can be presented as follows; 
in a quantum spatial search, the finding probability of a marked node should be amplified through the process where a  wavefunction spread uniformly in a network assembles at this target node. 
The majority of wavefunctions spread in the network are more likely to reach the target node simultaneously through paths with large length distribution, which allows the wavefunction to assemble most easily. 

This idea can be supported through an explicit relation between the probability amplitude and the path length distribution. 
Consider the probability amplitude for the quantum spatial search, given by $\pi \equiv \langle w | e^{- i \hat H t} | s \rangle$, where we divide the Hamiltonian 
$H = \hat H_L + \hat H_w$ as $H \equiv \hat H_A + \hat H_{\rm d}$ with $\hat H_A \equiv - \gamma \hat A$ and $ \hat H_{\rm d} \equiv - \sum_{i} \Gamma_i |i \rangle \langle i|$ for $\Gamma_i \equiv \gamma_w \delta_{i,w} - \gamma d_i$. 
By following the path-integral formalism with the Suzuki-Trotterization $\pi(t) = \lim\limits_{M\to \infty} \langle w | (e^{- i \hat H_A t /M } e^{- i \hat H_{\rm d} t /M } )^M | s \rangle$, the probability amplitude $\pi$ can be decomposed into a kernel ${\mathcal K}_M (i,j;t) \equiv \langle i |  e^{- i \hat H_A t /M } e^{- i \hat H_{\rm d} t /M }  | j \rangle $, which is reduced into 
\begin{align}
{\mathcal K}_M (i,j;t) = & 
    e^{i \Gamma_j t/M} 
    \sum\limits_{n=0}^\infty \frac{1}{n!} \left ( \frac{i \gamma t}{M} \right )^n
    f_n (i,j) 
\end{align} 
with $f_n (i,j) \equiv \langle i | \hat A^n | j \rangle$. 
As a result, the probability amplitude can be given by 
\begin{align}
   \pi(t) = & 
    \lim\limits_{M\to \infty} 
    \prod\limits_{m=1}^M 
    \sum\limits_{i_m=1}^N
    \sum\limits_{n_m=0}^\infty
    \frac{
    C_M^{(n_m, i_m)}(t) 
    }{\sqrt{N}} f_{n_m} (i_{m-1},i_{m})
    , 
\end{align}
where $C_M^{(n_m, i_m)}(t) \equiv e^{i \Gamma_{i_{m}} t/M} \left ( i \gamma t / M \right )^{n_m} /n_m !$. 
According to the path-integral formalism, 
the network emerges in each Trotter slice, and the hopping among adjacent nodes is realized between the adjacent Trotter slices, which is included through the function $f_{n_m} (i_{m-1},i_{m})$ for the node $i_{m-1}$ in the $(m-1)$-th Trotter slice and the node $i_m$ in the $m$-th Trotter slice. 
In particular, the function $f_n (i,j)$ for the unweighted graph just corresponds to the path length distribution of the path length $n$ from the node $j$ to $i$~\cite{Newmanbook}. 
As a result, the probability amplitude is related to the path length distribution of the network. 
Note that the probability amplitude can also be related to the path length distribution $P_n(i,j)$ by following the Green's function formalism with the propagator $G_{0i}(\omega) = 1/(\omega  - \gamma d_i + \gamma_w\delta_{i,w})$.

\begin{figure}[tbp]
    \centering
   \includegraphics[width=80mm]{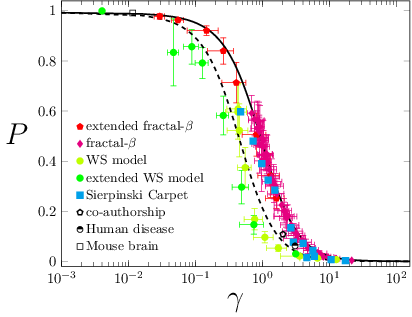}
   \caption{
    The correlation between the optimal hopping parameter $\gamma$ and the finding probability $P$ for unweighted graphs in the (extended) fractal-$\beta$ model, (extended) WS model, Sierpinski carpet, and networks of the co-authorship~\cite{newman2006finding}, human disease~\cite{goh2007human} and mouse brain~\cite{bigbrain}. The target node is chosen as a node with a minimum degree. 
    The solid and dashed lines are the fitting function (\ref{eq:scaling function}) for $D = 1$ and $D=2.11(1)$, respectively. 
    }
    \label{fig:scaling function}
\end{figure}

Finally, we report a relation between the finding probability $P$ and the optimal hopping parameter $\gamma$ for unweighted graphs in the WS model, the Sierpinski carpet, the (extended) fractal-$\beta$ model with the Sierpinski gasket (Fig.~\ref{fig:scaling function}). 
Those data are well fitted by the relation 
\begin{equation}
    P(\gamma)=\frac{1}{A+(D\gamma+B)^{C}}, 
    \label{eq:scaling function}
\end{equation}
with the fitting parameters $A = 0.991(1)$, $B = 0.086(5)$ and $C = 1.68(2)$. 
Here, $D$ is fixed as $D=1$ for the (extended) fractal-$\beta$ model and Sierpinski carpet (solid line in Fig.~\ref{fig:scaling function}), and $D$ is a fitting parameter, the value of which is $D = 2.11(1)$, for the (extended) WS model (dashed line in Fig.~\ref{fig:scaling function}). 
In the extended WS model and the extended fractal-$\beta$ model, which is based on the variant small-world model with shortcuts~\cite{Monasson1999,NEWMAN1999341,newman2003structure}, edges are added randomly with the probability $\beta$, which enables us to produce the data sets at $10^{-2} < \gamma < 10^0$. 
In the complete graph, an analytic form of $P (\gamma)$ is known~\cite{agliari2010quantum}, where $P(\gamma = 1/N)\simeq 1$~\cite{childs2004spatial}. 
The scaling function (\ref{eq:scaling function}) is also useful not only for the model network, but also for the real-world complex network~\cite{nr}, such as the co-authorship network~\cite{newman2006finding}, human disease network~\cite{goh2007human}, and mouse brain network~\cite{bigbrain} (Fig.~\ref{fig:scaling function}). 
We also found other classes of networks that do not show the correlation (\ref{eq:scaling function}): for example, 
the Erd\"os--R\'enyi random graph, hypercube lattices, fractal Cayley tree~\cite{agliari2010quantum}, T fractal network graph~\cite{agliari2010quantum}, and real-world networks such as dolphin network~\cite{lusseau2007evidence} and Facebook network~\cite{Traud:2011fs, traud2012social}. 


In summary, we propose the universal scaling hypothesis of the quantum spatial search in the complex networks. 
Using the continuous-time quantum walk in the Watts--Strogatz model, the fractal-$\beta$ model and the static scale-free network model, which covers the regular, random, small-world and scale-free networks, 
we have demonstrated that as a function of the average path length, the collapse plot can emerge in the finding probability, optimal time, as well as optimal hopping parameter. 
We also established the explicit relation between the probability amplitude of the target node and the path length distribution by using the path integral formalism. 
The idea of the complex network science is still useful in the quantum complex network systems, and our results provide a new insight into the connection between quantum physics and complex network science.

\section*{Acknowledgement}
S.W. was supported by JST, PRESTO Grant Number JPMJPR211A, Japan.
This work was supported by STADHI Satellite Lab, International Research Frontiers Initiative (Institute of Innovative Research, School of Environment and Society), Tokyo Institute of Technology.

\bibliographystyle{apsrev4-2}
\bibliography{preprint}

\end{document}


\preprint{APS/123-QED}

\title{Supplemental Materials for ``Universal scaling hypothesis of quantum spatial search in complex networks"}

\author{Rei Sato}
\email{1221706@ed.tus.ac.jp}
\affiliation{%
  Department of Physics, Tokyo University of Science, Shinjuku, Tokyo, 162-8601, Japan\\
}%
\affiliation{%
 Department of Transdisciplinary Science and Engineering, School of Environment and Society, Tokyo Institute of Technology,
2-12-1 O-okayama, Meguro-ku, Tokyo 152-8550, Japan\\
}%
\author{Tetsuro Nikuni}%
\affiliation{%
 Department of Physics, Tokyo University of Science,
 Shinjuku, Tokyo, 162-8601, Japan\\
}%

\author{Kayoko Nohara}%
\affiliation{%
 Department of Transdisciplinary Science and Engineering, School of Environment and Society, Tokyo Institute of Technology,
2-12-1 O-okayama, Meguro-ku, Tokyo 152-8550, Japan\\
}%

\author{Giorgio Salani}%
\affiliation{%
 Department of Transdisciplinary Science and Engineering, School of Environment and Society, Tokyo Institute of Technology,
2-12-1 O-okayama, Meguro-ku, Tokyo 152-8550, Japan\\
}%

\author{Shohei Watabe}%
\email{watabe@shibaura-it.ac.jp}
\affiliation{%
  Department of Physics, Tokyo University of Science, Shinjuku, Tokyo, 162-8601, Japan\\
}%
\affiliation{%
   Faculty of Engineering, Computer Science and Engineering, Shibaura Institute of Technology, Toyosu, Tokyo, 135-8548, Japan\\
}%


\maketitle

\appendix 

\section{Network Structure}\label{appndixI}

\begin{figure}[tbp]
    \centering
    \includegraphics[width=80mm]{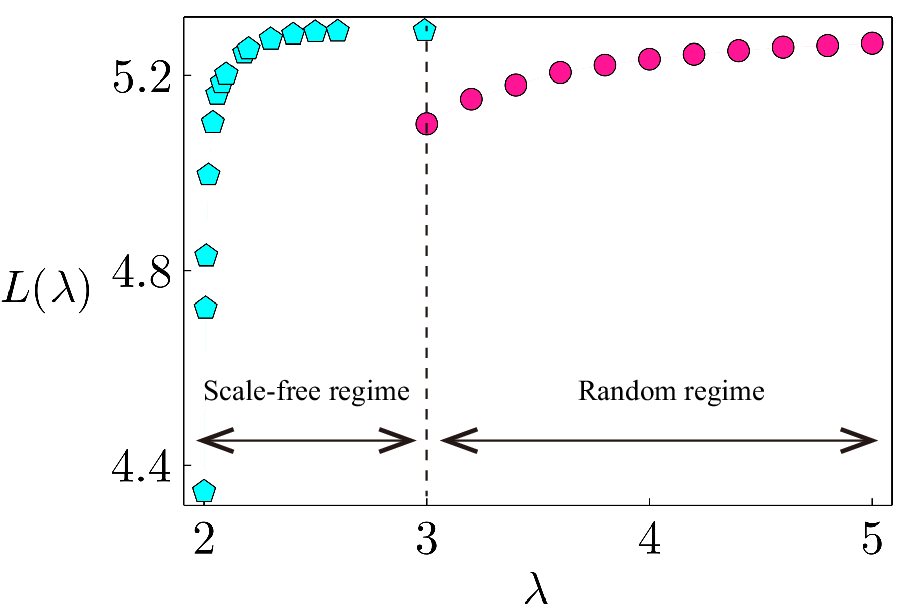}
    \caption{(Color online) The $\lambda$-dependence of the average path length $L$ in the static scale-free model, where $\lambda$ is an exponent of the degree distribution, given by $P_k\sim k^{-\lambda}$.} 
    \label{fig:path_scalefree}
\end{figure}

In this section, we summarize the complex network models used in this paper. 
For the small-world network, we employ the Watts--Strogatz (WS) model~\cite{watts1998collective}, the fractal-$\beta$ model~\cite{tomochi2010model}, which can cover not only the small-world network but also the regular ring, the regular fractal and random graphs. 
The method to generate a small-world network is proposed by Watts and Strogatz~\cite{watts1998collective}, which is based on a ring graph, but can be extended to other regular networks, such as a square, triangle and fractal lattices~\cite{tomochi2010model}. 
For example, starting with a regular ring network of $N$-vertices with the degree $k$, each node is randomly rewired with a probability $\beta$. 
By changing the probability from $\beta = 0$ to $\beta = 1$, this model interpolates the small-world network from the regular to random networks. 
For $\beta=0$, no edges are rewired, where the network keeps the original regular graph. 
For $\beta=1$, all the edges are randomly rewired, which results in a random network. 
The average path length is defined as $L=\frac{1}{N(N-1)}\sum\limits_{i,j}^{N}L_{ij}$, 
where $L_{ij}$ is the shortest path between nodes $i$ and $j$~\cite{barabasi2016network}. 
A local clustering coefficient $C_i$, which is used in the average clustering coefficient $C= \sum_{i=1}^{N}C_i / N$, is defined as $C_i = \frac{2}{k_i (k_i -1)} \Delta_i$ for a node $i$ with a degree $k_i$, and $\Delta_i$ represents the number of edges between $k_i$-neighbor nodes. 

For the scale-free network, we utilize the static scale-free network model~\cite{Goh2001}, because the number of nodes $N$ can be easily managed and constant through the network generation process rather than the growing network model, or 
the Barab\'{a}si-Albert model~\cite{balabasi1999}. 
To generate a scale-free network with a degree exponent $\lambda$, 
the following algorithm is used~\cite{Goh2001}; 
Each vertex is indexed by an integer $i$ for $i=1,\cdots,N$. 
A probability $p_i=i^{-\alpha}/\sum_{j=1}^{N}j^{-\alpha}$ is assigned to each vertex with a parameter $\alpha \in [0,1]$. 
Two different vertices $i$ and $j$ are randomly selected with probabilities $p_i$ and $p_j$, respectively, and if there are no links between them, a link is created. 
This process is stopped if the number of generated edges reaches a value $mN$ with a constant $m$, where the mean degree is given by $\langle k\rangle = 2m$. 
The exponent $\lambda$ in the degree distribution $P_k \sim k^{-\lambda}$ in this model can be related to the parameter $\alpha$, given by $\lambda=1+1/\alpha$~\cite{Goh2001}. 
Thus, by controlling the parameter $\alpha \in [0,1]$, we can generate the networks with the degree distribution for $2 \leq \lambda$~\cite{Goh2001}. 
For $\lambda<2$, the network is in an anomalous regime, where the average path length does not depend on $N$.  For $2<\lambda<3$, the network is in the scale-free regime with an ultra-small world property, where $L\sim\ln{\ln{N}}$. For $\lambda>3$, the network is in the random regime with a small-world property, where $L \sim\ln{N}$. The critical points emerge at $\lambda = 2$ with $L ={\rm const.}$ and at $\lambda = 3$ with $L =\ln{N}/\ln{\ln{N}}$ (Fig.~\ref{fig:path_scalefree}). 

For unweighted graphs, we employ elements of the adjacency matrix as $0$ or $1$, the value of which depends on the network structure we are interested in. 
For weighted graphs in this study, we replace elements with $1$ in the unweighted graph with the number randomly drawn from the uniform distribution over the interval $(0,1]$.


\section{Quantum Spatial Search}\label{appndixII}

\begin{figure}[tbp]
    \centering
    \includegraphics[width=130mm]{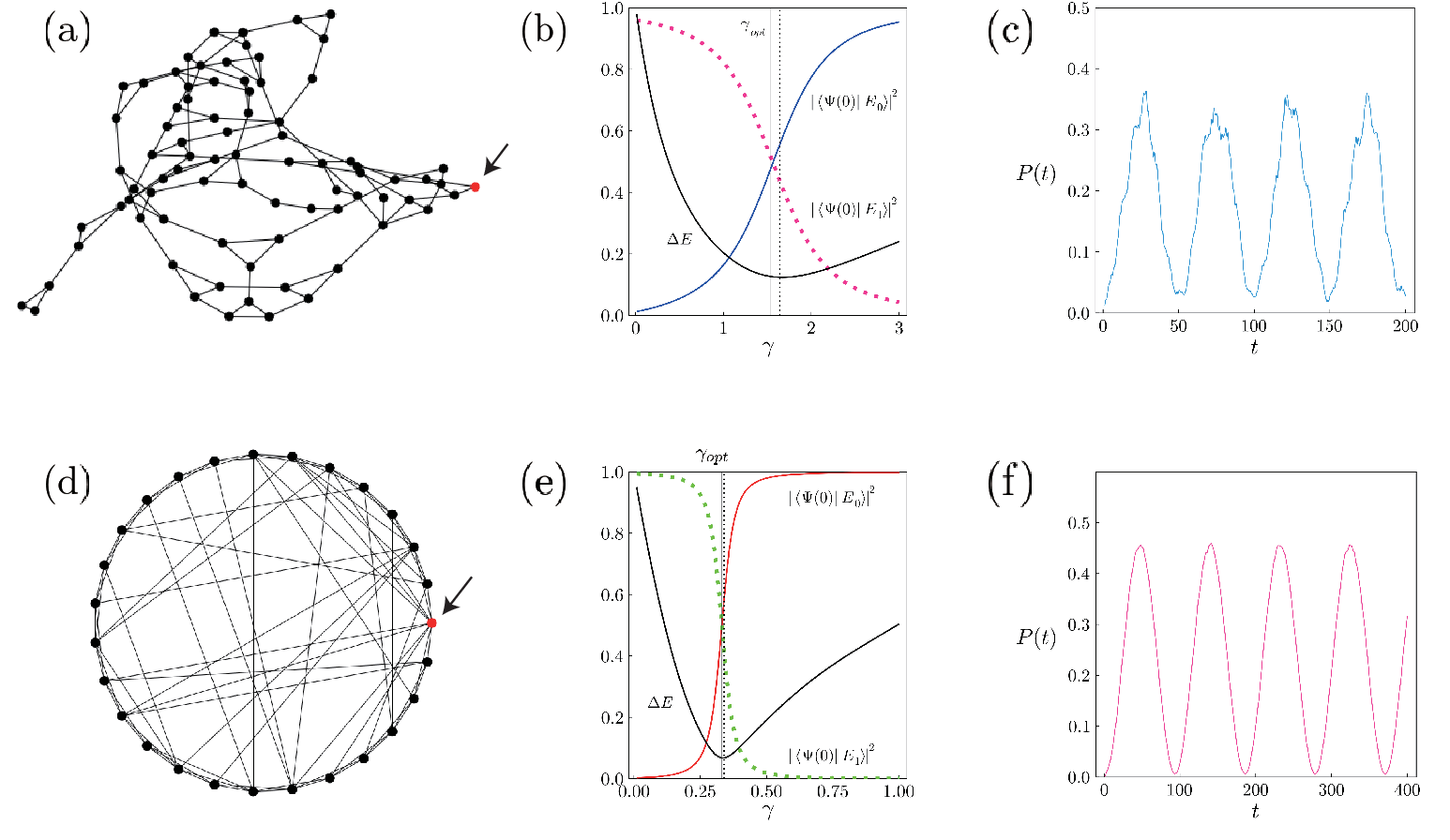}
    \caption{(Color online) (a)-(c) Fractal-$\beta$ model for $\beta=0.2$ and $N=3^4$ at the stage $S=4$, where the regular network at $\beta = 0$ is the Sierpinski gasket. 
    (d)-(f) The WS model for $\beta=0.6$. 
    (a) A network of a fractal-$\beta$ model, where the red point pointed by arrow is a target site $w$. 
    (b) Energy gap $\Delta E$ and overlaps $|\langle \Psi (0) | E_{0,1} \rangle |^2$ for the fractal-$\beta$ model. The dotted vertical line indicates $\gamma$ for the minimum energy gap. The solid vertical line at $\gamma=1.53$ is the position where $|\braket{\Psi(0)|E_0}|^2 = |\braket{\Psi(0)|E_1}|^2 \simeq 0.5$ holds. 
    (c) Time-dependence of the finding probability $P(t)$ at a target site $w$ with $\gamma_{} = 1.53$. 
    (d) A network of the WS model for $N=27$. 
    (e) Energy gap and overlaps for the WS model with $N=400$. 
    The dotted vertical line indicates $\gamma$ for the minimum energy gap. 
    The solid vertical line at $\gamma= 0.33$ is the position where $|\braket{\Psi(0)|E_0}|^2 = |\braket{\Psi(0)|E_1}|^2 \simeq 0.5$ holds. 
    (f) Time-dependence of the finding probability $P(t)$ at a target site $w$ with $\gamma = 0.33$ and $N=400$.}
    \label{fig:search_ws_fractal}
\end{figure}

The quantum spatial search for a target node $w$ using the continuous-time quantum walk is achieved by using the Schr\"odinger equation 
\begin{equation}
    i \frac{d}{dt}\ket{\Psi(t)} = \hat{H}\ket{\Psi(t)}, 
\end{equation}
where the time-independent Hamiltonian $\hat{H} = \hat{H}_L + \hat{H}_w$ is given by 
\begin{align}
    \hat{H}_L = & \gamma \hat{L}, 
    \\ 
    \hat{H}_w = & -\gamma_w |w \rangle \langle w| . 
\end{align} 
Here, $\gamma (>0)$ is a hopping parameter, and $\gamma_w (>0)$ is the strength of the target potential, which is often normalized as $\gamma_w = 1$. 
The Laplacian operator $\hat{L}$ is given by 
\begin{align}
    \hat L = \sum_{ij} L_{ij} |i\rangle \langle j| \equiv \hat D - \hat A, 
\end{align}
where the operators $\hat D$ and $\hat A$ are the degree and adjacency operators, respectively, given by  
\begin{align}
    \hat D =  \sum_{ij} D_{ij} |i\rangle \langle j|, \quad 
    \hat A =  \sum_{ij} A_{ij} |i\rangle \langle j|.   
\end{align}
Here, $D_{ij} \equiv d_i \delta_{ij}$ gives the degree matrix with the degree $d_i$ for the node $i$, and $A_{ij}$ is the adjacency matrix of the unweighted graph $G$ we are interested in, given by 
\begin{subequations}
\begin{align}
A_{ij}=\begin{cases}
1 & (i,j)\in G, \\
0 & \text{otherwise}.
\end{cases}
\end{align}
\end{subequations} 
For weighted graphs in this study, we replace $A_{ij} = 1$ in an unweighted graph with the number randomly drawn from the uniform distribution over the interval $(0,1]$.  

After generating networks such as the fractal-$\beta$ model (Fig.~\ref{fig:search_ws_fractal} (a)) and the WS model (Fig.~\ref{fig:search_ws_fractal} (d)), we optimize the hopping parameter $\gamma$, to estimate which we can choose two possible conditions~\cite{childs2004spatial}. The condition (i) is that the overlap between the ground state $|E_0\rangle$ of $\hat H$ and the initial equal-superposition state $|s\rangle$ is equal to that for the first-excited state $|E_1 \rangle$, which is close to $0.5$, {\rm i.e.,} $|\braket{s|E_0}|^2 = |\braket{s|E_1}|^2 \simeq 0.5$. The condition (ii) is that the energy gap $\Delta E$ between the first-excited and ground states is minimum. 
The earlier study~\cite{childs2004spatial} reported that the values of $\gamma$ in the conditions (i) and (ii) are almost equal. 

Both in the fractal-$\beta$ model (Fig.~\ref{fig:search_ws_fractal} (b)) and the WS model (Fig.~\ref{fig:search_ws_fractal} (e)), however, we found a small but significant difference between the values of $\gamma$ satisfying the overlap condition (i) and the minimum-energy-gap condition (ii). 
By comparing the finding probability $P(t)=|\bra{w}e^{-i\hat{H}t}\ket{s}|^2$ on a target node $w$ for two different values of $\gamma$, the higher peak and clearer periodicity emerge in the condition (i) rather than (ii) (Figs.~\ref{fig:search_ws_fractal} (c) and (f)). 
We thus estimate the optimal value $\gamma_{}$ by using the condition (i).


\section{Numerical results and data analysis}

\begin{table}[tbp]
    \centering
    \begin{tabular}{lcc} \hline\hline
               $t$ &  $10^4 \sim 10^6$ \\ 
               $\beta$ & $0$, $2.5\times10^{-3}, 7\times10^{-3}, 0.02, 0.04, 0.08, 0.2, 0.4, 0.6, 1$ \\ 
               $\lambda$ (scale-free) & $2.0, 2.007, 2.01, 2.02, 2.08, 2.99$  \\
               $\lambda$ (random)  & $3.0, 3.2, 3.4, 3.6, 3.8, 4.0, 4.2, 4.4$ \\
               \hline\hline
    \end{tabular}
    \caption{Parameter sets we used to find the collapse plot. $\beta$ is for the WS model and the fractal-$\beta$ model. $\lambda$ is for the static scale-free model.} 
    \label{tab:parameters}
\end{table}

In this section, we summarize details of numerical results and data analysis for unweighted graphs. 
In order to  analyze the scaling law, we perform numerical simulations, where the simulation time ranges from $t = 10000$ to $1000000$. 
The parameter sets for simulations are summarized in Table~\ref{tab:parameters}. 
We randomly generate networks by changing probabilities $\beta$ and $\lambda$.  

We determine the optimal value of the hopping parameter $\gamma$, the condition of which is given by $|\langle E_0 | s \rangle| = |\langle E_1 | s \rangle |$. 
Using the finding probability $P(\omega)$ in the frequency space after the Fourier transformation of $P(t)$, 
we extract the optimal frequency $\omega$ from the maximum value of $P(\omega)$, which determines the optimal time $Q$ that gives the peak-to-peak period of the finding probability. 
Using this optimal time $Q$, we gather the peak values of $P(t)$ in a long-time Schr\"odinger-dynamics simulation., which produces the average finding probability $P$.

\begin{figure}[tpb]
    \centering
    \includegraphics[width=150mm]{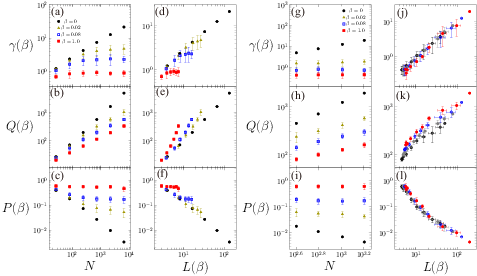}
    \caption{(Color online) 
    Quantum spatial search in the small-wold network. 
    (a)-(f) The fractal-$\beta$ model for the stage $S=3$-$8$, where the regular lattice at $\beta=0$ is the Sierpinski gasket. 
    (g)-(i) The WS model with $N=400$, $625$, $1024$, $1600$. 
    Here, we plot the data as a function of the number of nodes $N$ [(a)-(c), (g)-(i)], 
    and as a function of the average path length $L(\beta)$ [(d)-(f), (j)-(l)]. 
    The mean values of the optimal hopping parameter $\gamma$ [(a), (d), (g), (j)], the optimal time $Q$ [(b), (e), (h), (k)], 
    and the maximum finding probability $P$ at a target site [(c), (f), (i), (l)]. 
    }
    \label{fig:scaling_NL}
\end{figure}

In the following, we explain the trial-and-error process to find the collapse plot for the quantum spatial search in the complex network. 
We first analyzed the conventional scaling law for the fractal-$\beta$ and WS models as a function of the number of sites $N$, supposing the relation $\gamma(\beta)=\mathcal{O}(N^{a_{\gamma}})$, $Q(\beta)=\mathcal{O}(N^{a_{Q}})$ and $P(\beta)=\mathcal{O}(N^{a_{P}})$ (Figs.~\ref{fig:scaling_NL} (a)-(c) and (g)-(i)). 
However, we could not find characteristic universal behaviors in the WS model and fractal-$\beta$ model for various values of $\beta$. 

Since the path length is a key quantity in the complex network, we tried to plot the data as a function of the average path length $L(\beta)$ (Figs.~\ref{fig:scaling_NL} (d)-(f) and (j)-(l)). 
However, the simple plot as a function of the average path length $L(\beta)$ is also not useful to show the characteristic behavior, where the average path lengths spread in different orders. 
Therefore, the normalization of the path length is important for networks with the same number of nodes $N$ with different $\beta$, such as $L(\beta)/L(0)$.

\begin{figure}[tpb]
    \centering
    \includegraphics[width=180mm]{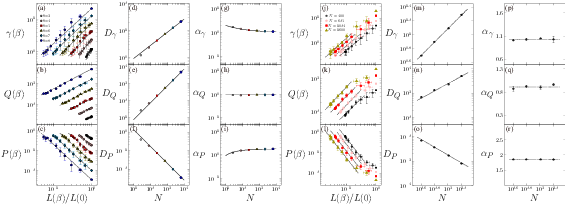}
    \caption{(Color online) 
    Scaling laws for the small-world network in the fractal-$\beta$ model (a)-(i) and in the WS model (j)-(r). 
    $\gamma$, $Q$ and $P$ as a function of the normalized average path length $L(\beta)/L(0)$ [(a)-(c), (j)-(l)]. 
    The dashed line is the fitting with $\gamma=D_{\gamma}\left(L(\beta)/L(0) \right)^{\alpha_{\gamma}}$, $Q=D_{Q}\left(L(\beta)/L(0)\right)^{\alpha_{Q}}$, $P=D_{P}\left(L(\beta)/L(0)\right)^{\alpha_{P}}$, respectively. 
    The scaling of the coefficients $D_{\gamma}$, $D_{Q}$ and $D_{P}$ as a function of the number of sites $N$ with the fitting form $uN^b$ [(d)-(f), (m)-(o)]. 
    The exponents $\alpha_{\gamma}, \alpha_{Q}, \alpha_{P}$ as a function of $N$ with the fitting form $p+qN^{r}$ [(g)-(i), (p)-(r)]. All fittings are performed for the stages $\rm S=5$-$8$ in the fractal-$\beta$ model.
    }
    \label{fig:path scaling}
\end{figure}

\begin{figure}[tbp]
    \centering
    \includegraphics[width=60mm]{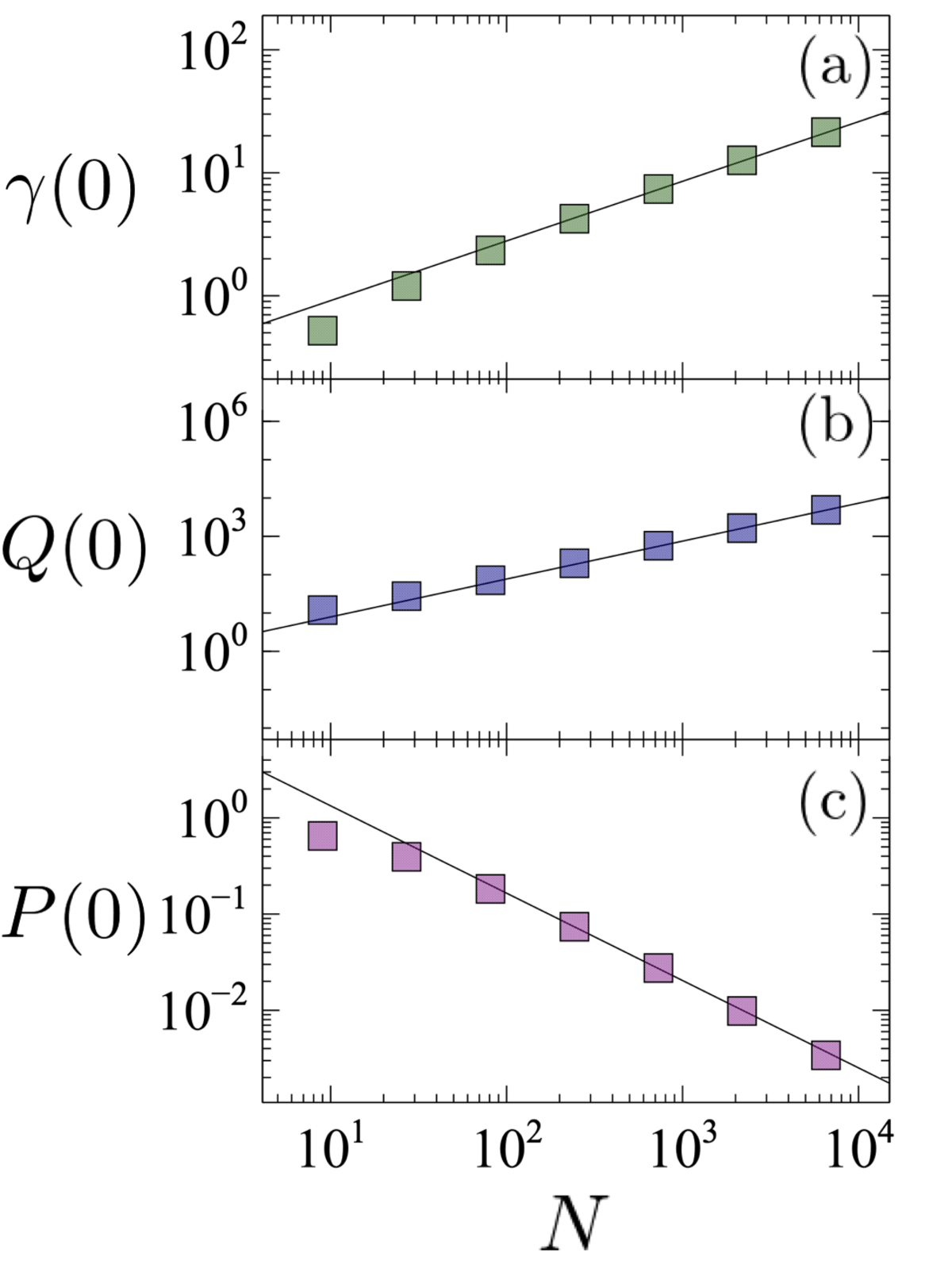}
    \caption{(Color online)
    Quantum spatial search on Sierpinski gaskets from ${\rm S}=2$ to $8$, 
    which corresponds to the fractal-$\beta$ model for $\beta = 0$. 
    (a) The hopping parameter $\gamma$, (b) the optimal time $Q$ and (c) the finding probability $P$. 
    Lines are fittings with the form $uN^b$. 
    All fittings are performed from $S=5$ to $8$.}
    \label{fig:search gasket}
\end{figure}

For the fractal-$\beta$ model, we analyze the data by supposing the relation $\gamma(\beta)=D_{\gamma}\left(L(\beta)/L(0)\right)^{\alpha_{\gamma}}$, $Q(\beta)=D_{Q}\left(L(\beta)/L(0)\right)^{\alpha_{Q}}$ and $P(\beta)=D_{P}\left(L(\beta)/L(0)\right)^{\alpha_{P}}$ (Figs.~\ref{fig:path scaling} (a), (d),(g)). 
Surprisingly, we find the universal exponent $\alpha_{\gamma,Q,P}$ for the normalized average path length $L(\beta)/L(0)$. 
We analyze the coefficients $D_{\gamma, Q, P}$ and the exponents $\alpha_{\gamma,Q,P}$ as a function of $N$ in the fractal-$\beta$ model (Figs.~\ref{fig:path scaling} (d)-(i)), which gives 
\begin{align}
    D_{\gamma}=0.299(9)N^{0.485(3)}, & \quad \alpha_{\gamma} = 1.04(3)+2.7(9)N^{-0.46(9)}, 
    \\ 
    D_{Q}=0.81(4)N^{0.989(6)}, & \quad \alpha_{Q} = 0.94(1)+0(2)N^{-0.6(8)},  
    \\ 
    D_{P}=10.8(5)N^{-0.908(9)}, & \quad \alpha_P = -1.82(3)+3.1(8)N^{-0.54(7)}. 
\end{align} 
We found that the scaling of the coefficients $D_{\gamma, Q, P}$ are excellently the same as that in the regular fractal network at $\beta =0$ (Fig.~\ref{fig:search gasket} and Table~\ref{tab:data of scaling relations}), i.e., 
\begin{align}
    D_{\gamma}=\gamma(0), \quad D_{Q}=Q(0), \quad D_{P}=P(0). 
\end{align} 
As a result, we have the scaling relations 
\begin{align}
\frac{\gamma(\beta)}{\gamma(0)} = \left(\frac{L(\beta)}{L(0)}\right)^{\alpha_\gamma}, 
\quad 
\frac{Q(\beta)}{Q(0)} = \left( \frac{L(\beta)}{L(0)}\right)^{\alpha_Q}, 
\quad 
\frac{P(\beta)}{P(0)} = \left( \frac{L(\beta)}{L(0)}\right)^{\alpha_P}. 
\label{eq:ws_scaling_law_P}
\end{align}
In the large-$N$ limit, the exponents converge as 
$\alpha_{\gamma} = 1.04(3)$, $\alpha_P = 0.94(1)$ and $\alpha_{Q} = -1.82(3)$, respectively. 
We can also successfully make the collapse plot in the WS model. 
Away from the regime for the regular network ($\beta = 0$), i.e., $L(\beta)/L(0) = 1$, 
the exponents for the WS model and the fractal-$\beta$ model are equivalent, the average values of which are given by 
\begin{align}
\alpha_\gamma = 1.08(6), \quad \alpha_Q = 0.90(5), \quad \alpha_P = -1.76(8). 
\end{align} 
We tried to make plots as a function of the average cluster coefficient $C(\beta)$ as well as the normalized average cluster coefficient $C(\beta)/C(0)$. However, we could not make any collapse plots of $\gamma$, $Q$ and $P$ for the (normalized) average cluster coefficients. 
In short, the scaling law on a small-world network is found to be determined by the average path length.

\begin{table}[tbp]
    \centering
    \begin{tabular}{ccccccc}\hline\hline
    & $D_\gamma$ & $D_Q$ & $D_P$ & $\gamma(0)$ & $Q(0)$ & $P(0)$ 
    \\ \hline
    u & $0.299(9)$ &  $0.81(4)$ &  $10.8(5)$ & $0.299(9)$ & $0.81(4)$ & $10.8(5)$ 
    \\ 
    b & $0.485(3)$ & $0.989(6)$ & $-0.908(9)$ & $0.485(3)$ & $0.989(6)$ & $-0.908(9)$
    \\ \hline\hline
    \end{tabular} 
    \caption{Scaling factors of $D_\gamma$, $D_Q$, $D_P$ and $\gamma(0)$, $Q(0)$, $P(0)$ in the fractal-$\beta$ model, where excellent agreements can be found. 
    We assume the scaling laws $uN^{b}$, where all fittings are performed from ${\rm S}=5$ to $8$.}
    \label{tab:data of scaling relations}
\end{table}

\begin{figure}[tbp]
    \centering
    \includegraphics[width=180mm]{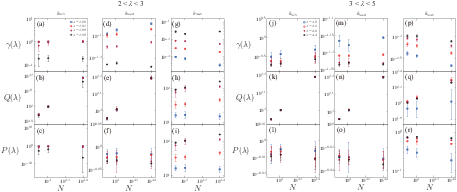}
    \caption{(Color online) 
    Quantum spatial search in the static scale-free model for the scale-free regime with the ultra-small world property (a)-(i) and for the random regime with the small-world property (j)-(r). 
    The mean values of the optimal hopping parameter $\gamma$ [(a), (d), (g), (j), (m), (p)], 
    the optimal time $Q$ [(b), (e), (h), (k), (n), (q)]
    and the maximum finding probability $P$ [(c), (f), (i), (l), (o), (r)] 
    as a function of the number of nodes $N$. 
    The target is selected as a node with the minimum degree $k_{\rm min}$ [(a)-(c), (j)-(l)], with the median degree $k_{\rm med}$  [(d)-(f), (m)-(o)] and with the maximum degree $k_{\rm max}$, the node with which is a hub [(g)-(i), (p)-(r)]. 
    }
    \label{fig:N_scaling_scalefree}
\end{figure}

\begin{figure}[tbp]
    \centering
    \includegraphics[width=180mm]{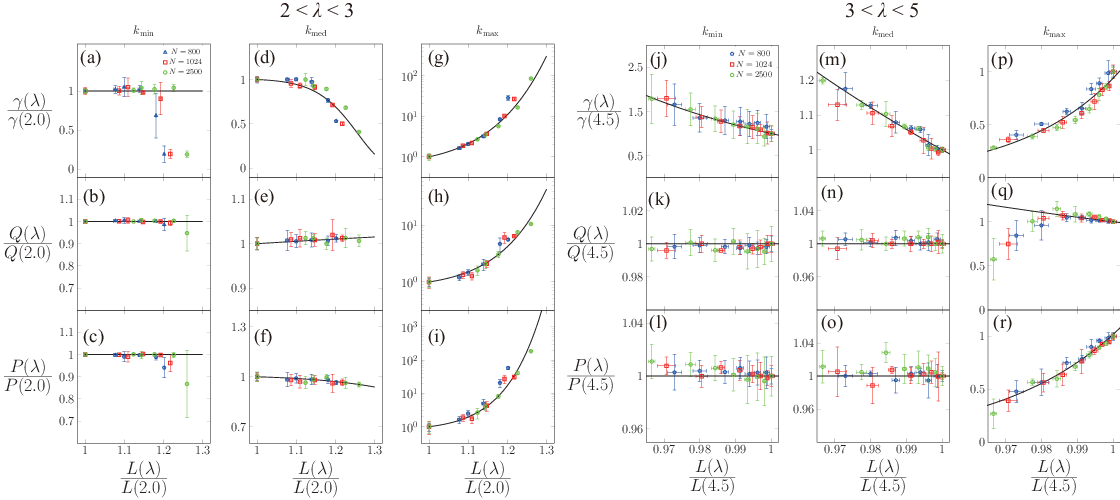}
    \caption{
    Collapse plots of data in Fig.~\ref{fig:N_scaling_scalefree} for the scale-free regime with the ultra-small world property (a)-(i) and for the random regime with the small-world property (j)-(r). 
    The mean values of the normalized optimal hopping parameter $\gamma (\lambda)/\gamma (\lambda_0)$ [(a), (d), (g), (j), (m), (p)], 
    the normalized optimal time $Q(\lambda)/Q(\lambda_0)$ [(b), (e), (h), (k), (n), (q)]
    and the normalized maximum finding probability $P(\lambda)/P(\lambda_0)$  [(c), (f), (i), (l), (o), (r)] 
    as a function of the normalized average path length $L(\lambda)/L(\lambda_0)$. 
    We chose $\lambda_0 = 2$ for $2 < \lambda < 3$ and $\lambda_0 = 4.5$ for $3 < \lambda$ as a reference. 
    The target is selected as a node with the minimum degree $k_{\rm min}$ [(a)-(c), (j)-(l)], with the median degree $k_{\rm med}$  [(d)-(f), (m)-(o)] and with the maximum degree $k_{\rm max}$, the node with which is a hub [(g)-(i), (p)-(r)]. 
    The solid lines are fittings with the forms summarized in Table~\ref{tab:scaling laws of scale-free network}. 
    }
    \label{fig:scaling23}
\end{figure}

We also confirmed that the collapse plots can be found in the scale-free networks for the normalized average path length.  
First of all, any characteristic universal behaviors cannot be found in the the $N$-dependence of $\gamma$, $Q$ and $P$ (Fig.~\ref{fig:N_scaling_scalefree}). 
In the simulation for the static scale-free model, we used $\lambda = \{2, 2.007, 2.01, 2.02, 2.08, 2.99\}$ for the scale-free regime with $N = \{ 800, 1024, 2500\}$, and $\lambda=\{ 3, 3.2, 3.4, 3.6, 3.8, 4, 4.2, 4.4\}$ for the random regime with $N=\{800, 1024, 2500\}$. 
We selected a target as a node with the minimum degree $k_{\rm min}$ [Figs.~\ref{fig:N_scaling_scalefree} (a)-(c), (j)-(l)], a node with the median degree $k_{\rm med}$ [Figs.~\ref{fig:N_scaling_scalefree} (d)-(f), (m)-(o)] and a hub, a node with the maximum degree $k_{\rm max}$ [Figs.~\ref{fig:N_scaling_scalefree} (g)-(i), (p)-(r)]. 

For normalized quantities $\gamma(\lambda)/\gamma(\lambda_0)$, $Q(\lambda)/Q(\lambda_0)$ and $P(\lambda)/P(\lambda_0)$, the collapse plots can be excellently organized as a function of the normalized average path length $L(\lambda)/L(\lambda_0)$, where data for different sizes of networks are well integrated into single curves (Fig.~\ref{fig:scaling23}). 
As a reference, we chose $\lambda_0 = 2$ for the scale-free regime ($2 < \lambda < 3$) and $\lambda_0 = 4.5$ for the random regime ($3 < \lambda$). 
The fitting functions we used are summarized in TABLE~\ref{tab:scaling laws of scale-free network}. 
As the parameter $\lambda$ approaches to $\lambda = 3$ from below, strong fluctuations emerge for targets with the minimum degree $k_{\rm min }$ (Figs.~\ref{fig:N_scaling_scalefree} (a)-(c)). 
Interestingly, a hub provides the clear path-length dependence for $\gamma(\lambda)/\gamma(\lambda_0)$, $Q(\lambda)/Q(\lambda_0)$ and $P(\lambda)/P(\lambda_0)$ (Figs.~\ref{fig:N_scaling_scalefree} (g)-(i), (p)-(r)). 
On the other hand, the normalized optimal time $Q(\lambda)/Q(\lambda_0)$ and the normalized finding probability $P(\lambda)/P(\lambda_0)$ do not show the clear path-length dependence and are almost constant for the cases of $k_{\rm min}$ (Figs.~\ref{fig:N_scaling_scalefree} (b), (c), (k), (l)), and with the median degree $k_{\rm med}$ (Figs.~\ref{fig:N_scaling_scalefree} (e), (f), (n), (o)).

\begin{table*}[tbp]
    \centering
    \begin{tabular}{rrrr}\hline\hline
    &$k_{\rm min}$ & $k_{\rm med}$ & $k_{\rm max}$\\\hline
     $\gamma(\lambda)/\gamma(2)$& $1$ &$\exp(-0.01(x^{19.9}-1))$&$0.94(2)x^{37(2)}$ \\
     $Q(\lambda)/Q(2)$& $1$ & $\exp(0.31(x^{0.18}-1))$ & $1.00(1)x^{-5(1)}$\\ 
     $P(\lambda)/P(2)$& $1$ & $\exp(-0.0091(x^{7.9}-1))$ & $1.02(1)x^{30(1)}$ \\\hline
    $\gamma(\lambda)/\gamma(4.5)$& $\exp(-15(x^{1.19}-1))$& $\exp(-10(x^{0.57}-1))$&$\exp{\left(4.4(4)(x^{9(1)}-1)\right)}$\\
     $Q(\lambda)/Q(4.5)$& $0.99813..$ & $1.002463..$ & $\exp{\left(-0.00019(7)(x^{-232(11)}-1)\right)}$\\ 
     $P(\lambda)/P(4.5)$& $1.00357..$ & $1.003576..$ & $\exp{\left(29(7)(x^{0.9(2)}-1)\right)}$\\ \hline\hline
    \end{tabular}
    \caption{Possible fittings for the collapse plots in the scale-free network shown in Fig.~\ref{fig:scaling23}, 
    where $x \equiv L(\lambda)/L(\lambda_0)$ with $\lambda_0 = 2$ for the scale-free regime (upper rows) and $\lambda_0 = 4.5$ for the random regime (lower rows), respectively. 
    In the upper rows for $k_{\rm min}$, we assumed the data as unity to ignore strong-fluctuation effects emerging around $\lambda = 3$. 
    }
    \label{tab:scaling laws of scale-free network}
\end{table*}

\section{Relation between the finding probability and the path length distribution}

\subsection{Path integral approach}

In this section, we explicitly derive the relation between the finding probability and the path length distribution. 
The finding probability of the quantum spatial search for a target site $| w \rangle$ is given by 
$|\pi (w,t)|^2$ 
with the probability amplitude $\pi (w, t)= \langle w | e^{- i \hat H t} | s \rangle$, 
where $|s\rangle$ is the equal-weight superposition state $|s \rangle = \sum_i | i \rangle /\sqrt{N}$ with the number of nodes $N$. 
Here $|i\rangle$ is the basis given for the node $i = 1, \cdots, N$. 

We divide the Hamiltonian $H = \hat H_L + \hat H_w$ as $H \equiv \hat H_A + \hat H_{\rm d}$, given by 
\begin{align}
    \hat H_A \equiv & \hat H_L - \gamma \hat D = - \gamma \hat A, \\
    \hat H_{\rm d} \equiv & \hat H_w + \gamma \hat D 
    = - \sum\limits_{i} \Gamma_i |i \rangle \langle i|, 
\end{align}
where $\Gamma_i = \gamma_w \delta_{i,w} - \gamma d_i$ and $\hat A$ is the adjacency operator. 
By using the Suzuki-Trotterization with the closure relation $\sum\limits_i |i \rangle \langle i| = \hat I$, the probability amplitude $\pi (w, t) = \langle w | e^{- i \hat H t } | s \rangle$ is given by 
\begin{align}
\pi(w, t) \equiv & \langle w | e^{- i \hat H t } | s \rangle 
\\ 
= & 
\lim\limits_{M\to \infty}
\langle w | (e^{- i \hat H_A t /M } e^{- i \hat H_{\rm d} t /M } )^M | s \rangle .
\\ 
= &  \lim\limits_{M\to \infty}
    \langle w | 
    e^{- i \hat H_A t /M } e^{- i \hat H_{\rm d} t /M } 
    \left (
    \sum\limits_{i_1= 1}^N | i_1 \rangle \langle i_1 |  
    \right )
    e^{- i \hat H_A t /M } e^{- i \hat H_{\rm d} t /M } 
    \left (
    \sum\limits_{i_2= 1}^N | i_2 \rangle \langle i_2 |  
    \right )
    \\
    & 
    \cdots
    \left ( 
    \sum\limits_{i_{M-1}= 1}^N | i_{M-1} \rangle \langle i_{M-1} |  
    \right )
    e^{- i \hat H_A t /M } e^{- i \hat H_{\rm d} t /M }
    \left ( 
    \sum\limits_{i_{M}= 1}^N | i_{M} \rangle \langle i_{M} |  
    \right )
    | s \rangle 
    \\ 
    = & \lim\limits_{M\to \infty}
    \sum\limits_{i_1= 1}^N 
    \langle w|  e^{- i \hat H_A t /M } e^{- i \hat H_{\rm d} t /M }  | i_1 \rangle 
    \cdot 
    \sum\limits_{i_2= 1}^N 
    \langle i_1 | e^{- i \hat H_A t /M } e^{- i \hat H_{\rm d} t /M } | i_2 \rangle
     \\ 
     &
    \cdots
    \sum\limits_{i_{M}= 1}^N 
    \langle i_{M-1} | e^{- i \hat H_A t /M } e^{- i \hat H_{\rm d} t /M } | i_M \rangle 
    \left ( 
    \frac{1}{\sqrt{N}}
    \sum\limits_{j=1}^N \langle i_M | j \rangle 
    \right ) 
    \\ 
    = & \frac{1}{\sqrt{N}} \lim\limits_{M\to \infty}
    \prod\limits_{m=1}^M 
    \left [
    \sum\limits_{i_m= 1}^N 
        {\mathcal K}_M ( i_{m-1},i_m;t)  
    \right ] , 
\end{align} 
where $| i_0 \rangle \equiv | w\rangle$ and the kernel is given by 
${\mathcal K}_M (i, j;t) \equiv 
    \langle i |  e^{- i \hat H_A t /M } e^{- i \hat H_{\rm d} t /M }  | j \rangle$. 
    
The kernel ${\mathcal K}_M (i, j;t)$ can be represented by using the path length distribution. 
Indeed, the kernel is reduced into 
\begin{align}
{\mathcal K}_M (i,j;t) = & 
    \langle i |  e^{- i \hat H_A t /M } e^{- i \hat H_{\rm d} t /M }  | j \rangle 
    \\ 
    = & 
    \langle i | 
    e^{- i \hat H_A t /M }
    \sum\limits_{n = 0}^\infty \frac{1}{n!} \left ( \frac{- i t}{M} \right )^n  \hat H_{\rm d}^n 
    | j \rangle 
    \\ 
    = & 
    \langle i | 
    e^{- i \hat H_A t /M }
    \sum\limits_{n = 0}^\infty \frac{1}{n!} \left ( \frac{- i t}{M} \right )^n  ( - \sum\limits_k \Gamma_k  | k \rangle \langle k |)^n 
    | j \rangle 
    \\ 
    = & 
    \langle i | 
    e^{- i \hat H_A t /M }
    \sum\limits_{n = 0}^\infty \frac{1}{n!} \left ( \frac{i t}{M} \right )^n  (  \sum\limits_k \Gamma_k^n  | k \rangle \langle k |)
    | j \rangle 
    \\ 
    = & 
    \sum\limits_{n = 0}^\infty 
\frac{1}{n!} \left ( \frac{i \Gamma_j t}{M} \right )^n 
    \langle i | e^{- i \hat H_A t /M } |j \rangle 
    \\ 
    = & 
    e^{i \Gamma_j t/M} 
    \langle i | e^{- i \hat H_A t /M } |j \rangle . 
\end{align}
The propagator $\langle i | e^{- i \hat H_A t /M } |j \rangle$ can be also reduced into 
\begin{align}
    \langle i | e^{- i \hat H_A t /M } |j \rangle = & 
    \langle i | \sum\limits_{n=0}^\infty \frac{1}{n!} \left ( - \frac{i t}{M} \right )^n \hat H_A^n | j \rangle 
    \\ 
    = & 
    \sum\limits_{n=0}^\infty \frac{1}{n!} \left ( \frac{i \gamma t}{M} \right )^n
    \langle i |  \hat A^n | j \rangle . 
\end{align} 
For unweighted graphs, the function $f_n (i,j) \equiv \langle i |  \hat A^n | j \rangle$ corresponds to the path length distribution with the adjacency matrix $\hat A$~\cite{Newmanbook}, which gives the number of the paths from the node $j$ to $i$ with the path length $n$. 

As a result, the kernel can be represented by the path length distribution as 
\begin{align}
    {\mathcal K}_M (i,j;t) = & 
    e^{i \Gamma_j t/M} 
    \sum\limits_{n=0}^\infty \frac{1}{n!} \left ( \frac{i \gamma t}{M} \right )^n
    f_n (i,j) , 
\end{align}
and the probability amplitude is thus given by 
\begin{align}
\pi(t) = & \frac{1}{\sqrt{N}} \lim\limits_{M\to \infty}
    \prod\limits_{m=1}^M 
    \left [
    \sum\limits_{i_m= 1}^N 
    \sum\limits_{n_m=0}^\infty 
    \frac{1}{n_m !} \left ( \frac{i \gamma t}{M} \right )^{n_m}
    e^{i \Gamma_{i_{m}} t/M} 
    f_{n_m} (i_{m-1},i_m) 
    \right ] . 
\end{align}

\subsection{Green's function approach}

The probability amplitude $\pi (w, t)$ is also represented by the Green's function. 
Here, we introduce the operator 
\begin{align}
	i\hat G(t,t')=e^{-i\hat H(t-t')}\theta(t-t'), 
\end{align}
which gives $iG_{ij}(t,t')=\braket{i|\hat G(t,t')|j}$. 
For $t>0$, the finding probability amplitude is given by 
\begin{align}
	\pi (w, t)  = \braket{w|e^{-i\hat H t}|s}=i\braket{w|\hat G(t)|s}= i\frac{1}{\sqrt{N}}\sum_i G_{wi}(t). 
\end{align}
Since the equation for $\hat G(t)$ is given by 
 \begin{align}
i\frac{d}{dt}\hat G(t)=\hat H\hat G(t)+\delta(t)\hat 1, 
 \end{align}
 the equation for $G_{ij}(t)$ is reduced into 
\begin{align}
	i\frac{d}{dt}G_{ij}(t)
	=\braket{i|i\frac{d}{dt}\hat G(t)|j}
=\braket{i|\hat H\hat G(t')|j}+\delta_{ij}\delta(t), 
\end{align}
which can be given by the matrix representation as 
\begin{align}
	i\frac{d}{dt}{\bf G}(t)={\bf H}{\bf G}(t)+{\bf I}\delta(t), 
\end{align}
where ${\bf G}$ and ${\bf H}$ are matrices the element of which are given by 
$G_{ij}$ and $H_{ij}$, respectively. Here, ${\bf I}$ is the identity. 

By using the Fourier transformation 
\begin{align}
	{\bf G}(t)=\int 
	\frac{d\omega}{2\pi}e^{-i\omega t}{\bf G}(\omega),~~
	{\bf G}(\omega)=\int dt \, e^{i\omega t}{\bf G}(t), 
\end{align}
we have the following equation 
\begin{align}
\int\frac{d\omega}{2\pi}\omega {\bf G}(\omega)e^{-i\omega t}
=\int\frac{d\omega}{2\pi}{\bf H}{\bf G}(\omega)e^{-i\omega t}
+{\bf I}\int\frac{d\omega}{2\pi}e^{-i\omega t}, 
\end{align}
which gives $\omega {\bf G}(\omega)={\bf H}{\bf G}(\omega)+{\bf I}$. 
As a result, we obtain 
\begin{align}
	{\bf G}(\omega)=(\omega{\bf I}-{\bf H})^{-1}, 
\end{align}
where 
${\bf H}=\gamma {\bf L} + {\bf H}_w = \gamma {\bf D} - \gamma {\bf A} + {\bf H}_w$,
with $H_{wij}=-\gamma_w \delta_{ij}\delta_{iw}$. 
The Green's function is now given by 
\begin{align}
	{\bf G}(\omega)= & 
	(\omega{\bf I} - {\bf H}_w - \gamma {\bf D} + \gamma {\bf A})^{-1}
\\ 
= & [{\bf I}-\gamma {\bf G}_0(\omega){\bf A}]^{-1}{\bf G}_0(\omega),
\end{align}
where the non-perturbative Green's function 
\begin{align}
{\bf G}_0(\omega)=(\omega{\bf I}-{\bf H}_w - \gamma {\bf D})^{-1} 
\end{align}
is the diagonal matrix, the element of which is given by 
\begin{align}
	G_{0i}(\omega) \equiv & G_{0ii} (\omega) = \frac{1}{\omega  - \gamma d_i + \gamma_w\delta_{i,w}} . 
\end{align} 
By expanding the term $[{\bf I}-\gamma{\bf G}_0(\omega){\bf A}]^{-1}$, 
the Green's function is given by 
\begin{align}
	{\bf G}(\omega)
	= & 
	{\bf G}_0(\omega)
	+\gamma {\bf G}_0(\omega){\bf A}{\bf G}_0(\omega)
	+\gamma^2 {\bf G}_0(\omega){\bf A}{\bf G}_0(\omega)
	+
	\cdots
	\\ 
	= & \sum_{n=0}^{\infty}[\gamma {\bf G}_0(\omega){\bf A}]^n
	{\bf G}_0(\omega). 
\end{align}
The kernel for the quantum spatial search can be given by 
\begin{align}
	K(\omega) \equiv & 
	\sum_{i}G_{wi}(\omega)
	= \sum_i  \sum_{n=0}^{\infty}\gamma^n [({\bf G}_0(\omega){\bf A})^n
	{\bf G}_0(\omega)]_{wi}. 
	\label{K}
\end{align}
Here, we have the relation 
\begin{align}
\{[{\bf G}_0(\omega){\bf A}]^n
	{\bf G}_0(\omega)\}_{wi}
	= & \{{\bf G}_0(\omega){\bf A}[{\bf G}_0(\omega){\bf A}]^{n-1}
	{\bf G}_0(\omega)\}_{wi} 
	\\ 
	= & G_{0w}(\omega)\{[{\bf A}{\bf G}_0(\omega)]^n\}_{wi}
\end{align}

For simplicity, we assume that all the degrees for $i\neq w$ is $d_{i (\neq w)}=d$, which provides 
\begin{align}
	G_{0i}(\omega)=\frac{1}{\omega+\gamma_w - \gamma d_w}\delta_{iw}+
	\frac{1}{\omega - \gamma d}(1-\delta_{iw})
	=G_{0w}(\omega)\delta_{iw}+G'(\omega)(1-\delta_{iw}), 
\end{align}
where $G' \equiv G_{0x(\neq w)} = 1/(\omega - \gamma d)$.  
Using this relation, we have 
\begin{align}
	\{[{\bf A}{\bf G}_0(\omega)]^n\}_{ij}
	= & \sum_{i_1}\cdots \sum_{i_{n-1}}
	A_{i i_1}G_{0 i_1}A_{i_1i_2}G_{0i_1}
	\cdots A_{i_{n-2}i_{n-1}}G_{0i_{n-1}}A_{i_{n-1}j}G_{0j} 
\\ 
= & [G'(\omega)]^{n-1}G_{0x}(\omega)	{\sum_{i_1}}'\cdots {\sum_{i_{n-1}}}'
A_{ii_1}A_{i_1i_2}
\cdots A_{i_{n-2}i_{n-1}}A_{i_{n-1}j}\nonumber\\
&+G_{0w}(\omega)[G'(\omega)]^{n-2}G_{0i}(\omega)
	\Biggl\{{	A_{iw}\sum_{i_2}}'\cdots {\sum_{i_{n-1}}}'
	A_{wi_2}\cdots A_{i_{n-2}i_{n-1}}A_{i_{n-1}j}
\nonumber\\ &
	+\left({\sum_{i_1}}'A_{ii_1}
	A_{i_1w}\right)
	{\sum_{i_3}}'\cdots {\sum_{i_{n-1}}}'
	A_{wi_{3}}\cdots A_{i_{n-1}j}+\cdots\Biggr\}\nonumber\\
	&+[G_{0w}(\omega)]^2[G'(\omega)]^{n-3}G_{0i}(\omega)\biggl\{\cdots\biggr\}+\cdots, 
\end{align} 
where we used the notation ${\sum_{i}}'=\sum_{i}(1-\delta_{iw})$, which represents the sum without $i =w$. 

Suppose an unweighted graph, and let $P^{k}_n(i,j)$ be the path length distribution with a path length $n$ starting from $j$ to $i$ passing through the node $w$ $k$-times. 
In particular, the path length distribution for the path never through the node $w$ is given by 
\begin{align}
P_n^{(0)}(i,j)={\sum_{i_1}}'\cdots {\sum_{i_{n-1}}}'
A_{ii_1}A_{i_1i_2}
	\cdots A_{i_{n-2}i_{n-1}}A_{i_{n-1}j} . 
\end{align}
Using this relation, we have 
\begin{align}
	\{[{\bf A}{\bf G}_0(\omega)]^n\}_{ij}
	=&[G'(\omega)]^{n-1}G_{0y}(\omega)P_n^{(0)}(i,j)	
	\nonumber\\
&	+G_{0w}(\omega)[G'(\omega)]^{n-2}G_{0j}(\omega)
\sum_m P_m^{(0)}(i,w)
P_{n-m}^{(0)}(w,j)\nonumber\\
&+[G_{0w}(\omega)]^2[G'(\omega)]^{n-3}G_{0j}(\omega)
\sum_{m}\sum_lP^{(0)}_m(i,w)P^{(0)}_l(w,w)
P^{(0)}_{n-m-l}(w,j) \cdots
\\
 = & 
 \sum_k[G_{0w}(\omega)]^k[G'(\omega)]^{n-k-1}G_{0j}(\omega)P^{(k)}_n(i,j), 
\end{align}
where we used 
$P^{(k+1)}_n(i,j) = \sum_m P^{(0)}_m(i,w)P_{n-m}^{(k)}(w,j)$. 
Therefore, we have 
\begin{align}
	K(\omega)
	=\sum_{n=0}^{\infty}\gamma^n
	\sum_k[G_{0w}(\omega)]^k[G'(\omega)]^{n-k-1}\sum_jG_{0j}(\omega)P^{(k)}_n(w,j). 
\end{align} 
In short, the finding probability amplitude can be also represented by using the Green's function and the path length distribution.

\section{Relations between the optimal finding probability $P$ and the optimal hopping parameter $\gamma$}

\begin{table}[tbp]
    \centering
\begin{tabular}{ccccc} 
    Square lattice & & Hexagonal lattice & &  Hypercubic lattice  
    \\ 
    \begin{tabular}{ccc} \hline\hline
    $N$ & $400$ & $900$ \\ \hline
    $\gamma$ & $1.66$ & $1.96$ \\ 
    $P$ & $0.141$ & $0.0760$ \\ \hline\hline
    \end{tabular} 
    &   \hspace{10pt} & 
    \begin{tabular}{c|cccc|cc|cc} \hline\hline 
    $N$ & $400$ & $900$ \\ \hline
    $\gamma$ & $3.09$ & $3.53$ \\ 
    $P$ & $0.0698$ & $0.0423$ \\ \hline\hline
    \end{tabular} 
    &   \hspace{15pt} & 
    \begin{tabular}{ccccc} \hline\hline
    $d$ & $8$ & $9$ & $10$ & $11$ \\  
    $N$ & $2^8$ & $2^9$ & $2^{10}$ & $2^{11}$ \\  \hline 
    $\gamma$ & $0.161$ & $0.141$ & $0.114$ & $0.104$ \\ 
    $P$ & $0.544$ & $0.423$ & $0.800$ & $0.767$ \\ \hline\hline
    \end{tabular} 
    \end{tabular}
     \\\vspace{15pt}   
    Erd\"os--Renyi model \\ 
    \begin{tabular}{c|ccc|ccc}\hline\hline
    $N$ & $400$ & & & $900$ & & \\ 
    $p$ & $0.1$ & $0.4$ & $1.0$ & $0.1$ & $0.4$ & $1.0$ \\ \hline 
    $\gamma$ & $0.0273$ & $0.00716$ & $0.00310$ & $0.0122$ & $0.00410$ & $0.00210$ \\ 
    $P$  & $0.871$ & $0.363$ & $0.178$ & $0.461$ & $0.0297$ & $0.0114$ \\ \hline\hline
    \end{tabular}
    \\\vspace{10pt}
    SC$(s,s-2)$ \\ 
    \begin{tabular}{c|cccc|ccc|ccc|cccccccccccc}  \hline\hline
    $s$ & $3$ &  &  &  & $4$ &  &  & $5$ &  &  & $6$ & & \\  
    $N$ & $8^2$ & $8^3$ & $8^4$ & $8^5$ & $12^2$ & $12^3$ & $12^4$ & $16^2$ & $16^2$ & $16^2$ & $20^2$ & $20^3$ & $20^4$\\  \hline 
    $\gamma$ & $0.470$ & $1.52$ & $2.88$ & $4.57$ & $0.72$ & $2.644$ & $5.92$ & $0.97$ & $4.01$ & $10.5$ & $1.21$ & $5.61$ & $17.25$ \\ 
    $P$ & $0.597$ & $0.284$ & $0.0764$ & $0.0153$ & $0.48$ & $0.134$ & $0.0201$ & $0.390$ & $0.0720$ & $0.00670$ & $0.324$ & $0.0433$ & $0.00284$\\\hline\hline
    \end{tabular}
    \\\vspace{10pt}
    Real complex networks \\ 
    \begin{tabular}{cccccc}\hline\hline
    &   Dolphin\cite{lusseau2007evidence} & Mouse brain\cite{bigbrain} & Co-authorship\cite{newman2006finding} & Human disease\cite{goh2007human} & Facebook\cite{Traud:2011fs, traud2012social}  
    \\ \hline \hline 
    $N$ & $62$ & $213$ & $379$ & $516$ & $1446$ \\  \hline 
    $\gamma$ & $1.67$ & $0.0115$ & $2.01$ & $3.05$ & $2.22$ \\ 
    $P$ & $0.661$ &  $0.990$ & $0.108$ & $0.0617$ & $0.758$\\\hline\hline
    \end{tabular}
    
    \caption{
    Data sets for discussing the relation between the finding probability $P$ and the optimal hopping parameter $\gamma$, which is used in Fig. 3 in the main body. 
    SC($s,s-2$) represents the Sierpinski carpet with the scaling factor $s$. The target node is chosen as a node with a minimum degree. }
    \label{tab:data for scaling function}
\end{table}

We here summarize the relations between the optimal hopping parameter $\gamma$ and the finding probability $P$. 
Note that the function $P (t, \gamma)$ was analytically derived for the complete graph~\cite{agliari2010quantum}. 
We extend this discussion to the complex networks by employing unweighted graphs in the WS model, the Sierpinski carpet, the (extended) fractal-$\beta$ model with the Sierpinski gasket (See Fig. 3 in the main body). 
In order to produce the data sets for $10^{-2} < \gamma < 10^0$, 
we extend the WS model and fractal-$\beta$ models by introducing shortcuts~\cite{Monasson1999,NEWMAN1999341,newman2003structure}, where edges are added randomly with the probability $\beta$. 
The network is regular at $\beta=0$ and becomes the complete graph at $\beta = 1$ when the number of added shortcuts are sufficient. 
To analyze the data for the Sierpinski carpet and the (extended) fractal-$\beta$ model with the Sierpinski gasket, we assumed three fitting functions: 
\begin{align}
    f_1(\gamma) \equiv & \frac{1}{A+(D \gamma+B)^{C}}, 
    \label{eq:scaling function}
    \\ 
    f_2(\gamma) \equiv & \frac{1}{A^{\gamma}}, 
    \\ 
    f_3(\gamma) \equiv & \frac{1}{A\gamma^{B}+C \gamma} \tanh{\left (\sqrt{\frac{\gamma}{\gamma+D}} \right )}, 
\end{align}
where $A$, $B$, $C$ and $D$ are fitting parameters. 
Here, the parameter $D$ in $f_1$ is temporarily fixed as $D=1$. 
We evaluated the accuracy of these fitting functions by using the root mean squared percentage error for the fitting function $f_{i=1,2,3}$, given by 
\begin{align}
    {\rm RMSPE}_i = \sqrt{\frac{1}{N} \sum_{j=1}^{N}
    \left (
        \frac{ P_j - f_i  }{ P_j } 
    \right )^2
    }, 
\end{align}
where $P_j$ is the simulation data of the finding probability, the total number of which is $N$. 
Results of our simulations are 
${\rm RMSPE}_1 = 0.1423$, ${\rm RMSPE}_2 = 0.2969$ and ${\rm RMSPE}_3 = 0.40178$. 
Therefore, we propose the scaling function $f_1$ 
with the fitting parameters $A = 0.991(1)$, $B = 0.086(5)$ and $C = 1.68(2)$. 
The complete graph gives the relation $P(\gamma = 1/N) \simeq 1$~\cite{childs2004spatial}, which is the scope of the scaling function $f_1$ in the large-$N$ regime. 

The scaling function of the (extended) WS model can be given by a variant of $f_1$, 
where the parameter $D$ is now turn to be a fitting parameter, which gives $D = 2.11(1)$. Here, the values of $A$, $B$ and $C$ are the same as those for the Sierpinski carpet and the (extended) fractal-$\beta$ model. 
The scaling function is also useful not only for the theoretical model such as the (extended) fractal-$\beta$ model including the complete graph, but also real-world complex networks found in the open data~\cite{nr}, such as the co-authorship network~\cite{newman2006finding}, human disease network~\cite{goh2007human} and mouse brain network~\cite{bigbrain}. 

We also find network classes where the relation in Eq.~(\ref{eq:scaling function}) does not work, which are, for example, the Erd\"os--Renyi random graph, hypercube lattices, fractal Cayley tree~\cite{agliari2010quantum}, T fractal network graph~\cite{agliari2010quantum}, and real complex networks such as dolphin network~\cite{lusseau2007evidence} and Facebook network~\cite{Traud:2011fs, traud2012social}. The square and hexagonal (honeycomb) lattices are marginal. 
An interesting expectation is that the difference between two classes may depend on the spectral dimensions. 
A class of the networks that follows the relation in Eq. (\ref{eq:scaling function}) may have the spectral dimension $d_{\rm s} < 2$, whereas other graphs may have the spectral dimensions $d_{\rm s} > 2$, where the behavior of the quantum spatial search is known to change around the critical spectral dimension $d_{\rm s} = 2$~\cite{10.7566/jpsj.87.085003,PhysRevA.86.012332,sato2020scaling}.


\bibliographystyle{apsrev4-2}
\bibliography{preprint.bib}